\documentclass{ws-procs9x6}

\begin{document}

\title{Wavelet transform and diffusion equations:
applications to the processing of the ``Cassini'' spacecraft
observations}

\author{E.B. Postnikov}

\address{Kursk State University, Theoretical Physics Department\\
305000, Radishcheva st., 33, Kursk, Russia\\
E-mail: postnicov@mail.ru}

\author{A.Loskutov}

\address{M.V. Lomonosov Moscow State University, Physics Faculty, \\
119899, Leninskye Gory, GSP-2, Moscow, Russia\\
E-mail: loskutov@polly.phys.msu.ru}

\maketitle

\abstracts{We show that continuous transform with the complex
Morlet wavelet is easily performed if we replace the integration
of the fast-oscillation function by the solution of the diffusion
differential equations. The most important advantage of this
approach is that the initial data can be represented by
non-uniform sample of an arbitrary node number. We apply the
proposed method to the processing of the image of the Saturn
A-ring obtained from the Cassini spacecraft. Also we have got via
the wavelet transform using PDE, the local correlation coefficient
of the signal and the harmonic function.}

\section{Introduction}

As is known, during last 20 years the wavelet transform is one of
the contemporary methods for the signal processing. The main
advantage of this approach is the high localization of the base
functions in both, spatial and frequent regions (see, e.g.,
\cite{Poularikas}, \cite{Dobechi}). This allows us to analyze
effectively non-stationary signals and signals with certain
singularities. For our convenience, the continuous wavelet
transform
\begin{equation}
\label{CWT} w(a,b) = \int\limits_{ - \infty }^{ + \infty }
{f(t)\psi^* \left( {\frac{{t - b}}{a}} \right)\frac{{dt}}{a}},
\end{equation}
(where the asterisk means the complex conjugation) with the
amplitude norm
\begin{equation}
\label{norm} \int_{-\infty}^{\infty}\left| \psi \left( {\frac{{t -
b}}{a}} \right)\right|\frac{dt}{a}={\rm const}.
\end{equation}
is the most convenience.

The wavelet basis which is often used for the signal processing is
the complex Morlet basis. Its exact form that satisfy the
admissible condition
$$
\int_{-\infty}^{\infty}\psi(\xi)d\xi=0,
$$
has the following form \cite{Dobechi}:
\begin{equation}
\label{MorletExact}
 \psi(\xi ) =
 \frac{1}{\sqrt{2\pi}} \left(
e^{ - i\omega _0 \xi } -e^{-\frac{\omega^2_0}{2}} \right)
 e^{ - \frac{\xi^2}{2}}.
\end{equation}
In this case, the corresponding wavelet-transform $w(a,b)$ plays
the role of the local spectrum distribution by period $a$ in the
neighborhood of the point $b$. This feature allows, on the basis
of the instant frequency notion \cite{Vela-Arevalo}, to develop
the method of the analysis of the solution for Hamiltonian
systems. Another approach developed on the basis of the
mean-square deviation of the investigated series from the model
data (but without of the wavelets) has been proposed in
\cite{Loskutov}. In the present paper, we show that these methods
are similar to each other.

It should be however noted, that the standard realization of the
continuous wavelet transform based on the transition to the
frequency domain and fast Fourier transform (FFT) has certain
disadvantages. They follow directly from FFT: the initial data
should be presented by the sample $2^N$ equispaced nodes. Neglect
of this condition leads to the algorithm complication and the lost
of the accuracy. But the data sets obtained via dynamical systems
and some astronomical observations can be presented by
sufficiently non-equispaced sample.

Thus, we should rely upon an alternative method of the evaluation
of the wavelet transform with the real-valued Gaussian family:
$$
\psi_n (\xi ) =   \frac{1}{{\sqrt {2\pi } }}\frac{{d^n }}{{d\xi ^n
}}e^{ - \frac{{\xi ^2 }}{2}}, \quad n\geq 1,
$$
This approach presents the wavelet-transform as a derivative of
the diffusively smooth initial function \cite{Mallat}:
$$
w(a,b)=a^n\frac{\partial^n}{\partial b^n}
\int_{-\infty}^{+\infty}f(t)\psi_0\left( {\frac{{t - b}}{a}}
\right)\frac{dt}{a}.
$$
This function can be found as a solution of the diffusion equation
with the iterative or finite-difference methods. In this case,
there are no restrictions to the initial date representation. In
addition, this approach can be adapted for the multidimensional
systems. For example, in \cite{Kestener} it successfully used for
the analysis of $3D$ turbulent solutions.

The main purpose of the present paper is to reduce the continuous
integral wavelet transform with complex wavelet to the solution of
the partial differential equations (a Cauchy problem). It allows
us to use the developed algorithm in the local frequency analysis.
This approach (which cannot be realized by the standard FFT)
additionally makes it possible almost arbitrary scale
decompositions. This problem is very important now in application
to small-scale structure of the planetary rings, in particular,
the Saturn rings (see, e.g. a review \cite{Esposito}). For the
first time, analysis the Voyager-2 data by the wavelet transform
has been undertaken in \cite{Bendjoya}. The authors using the real
wavelets, detect the structure of the rings in the noisy images.
At the same time, the problem related to the local periodicity is
still open, but it can be resolved by the complex wavelet
transform. This is due to the fact that in the present the
necessary data are available by the Huygens-Cassini mission.

\section{Continuous transform with the Morlet wavelet}

For applications, when $\omega_0$ is a large enough, in
(\ref{MorletExact}) one can neglect the second summand. Thus, in
the present Part we will used the following simplification:
\begin{equation}
\label{Morlet}
 \psi(\xi ) =
\frac{e^{{\frac{\omega_0^2}{2}}}}{{\sqrt {2\pi } }}e^{ - i\omega
_0 \xi } e^{ - \frac{{\xi ^2 }}{2}}.
\end{equation}
This expression corresponds to the normalization (\ref{norm}) with
${\rm const}=\exp(\omega _0^2/2)$. Let us show that such a choice
(owing to the violation of admissible conditions) makes it
possible to use the transformed function as an initial condition
for the equation
\begin{equation}
\label{PDE}
    \left( {a\frac{{\partial ^2 }}{{\partial b^2 }} - \frac{\partial
}{{\partial a}} - i\omega _0 \frac{\partial }{{\partial b}}}
\right)w(a,b) = 0
\end{equation}
which is satisfied by the wavelet-image (\ref{CWT}) with the
Morlet wavelet (see \cite{Haase}). Let us write the wavelet-image
as a sum of the real and the imaginary parts
$$
w(a,b)=u(a,b) + iv(a,b).
$$
With respect to these variables Eqs.(\ref{PDE}) may be presented
by the following system:
\begin{eqnarray}
\label{systemU}
   {\frac{{\partial u}}{{\partial a}} = a\frac{{\partial ^2 u}}{{\partial b^2 }} +
\omega _0 \frac{{\partial v}}{{\partial b}}}\\
 \label{systemV}  {\frac{{\partial v}}{{\partial a}} = a\frac{{\partial ^2 v}}{{\partial b^2 }} -
\omega _0 \frac{{\partial u}}{{\partial b}}}.
\end{eqnarray}
To find the corresponding initial conditions let us rewrite the
continuous transform (\ref{CWT}) with the kernel (\ref{Morlet}) in
the form:
\begin{equation}
\label{integral}
 w(a,b) =
\int\limits_{ - \infty }^{ + \infty } {f(t)\frac{{e^{ -
{\textstyle{1 \over 2}}\left( {{\textstyle{{t - b} \over a}} -
i\omega _0 } \right)^2 } }}{{\sqrt {2\pi a^2 } }}dt}.
\end{equation}
As is known, the integral (\ref{integral}) does not depends on the
imagine subtrahend in the exponent. Also, in the limit $a \to 0$
the transformation kernel is a delta-function. Therefore, we can
get initial conditions
$$
\begin{array}{l}
 u(0,b) = Re(f(b)), \\
 v(0,b) = Im(f(b)) \\
 \end{array}
$$
for the system of differential equations
(\ref{systemU})--(\ref{systemV}).

\begin{figure}
\begin{center}
\includegraphics[scale=0.5]{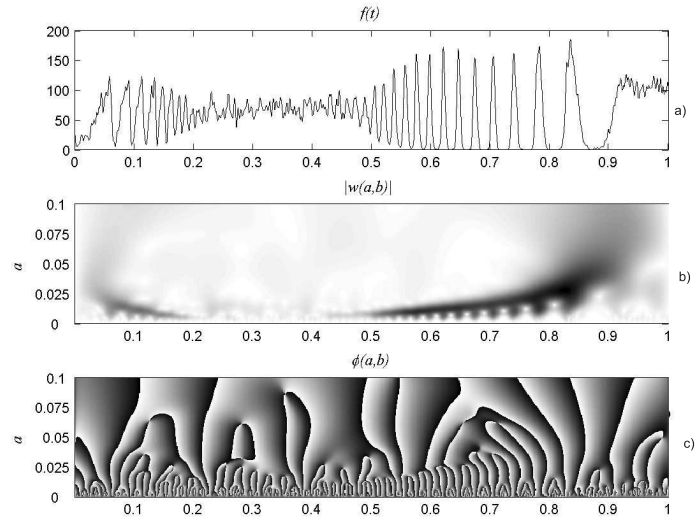}
\end{center}
\caption{The radial distribution of particle, modulus and the
phase of the wavelet-transform.} \label{CWTfig}
\end{figure}

Let us apply the described algorithm to the analysis of radial
distributions of the matter density in the Saturn A-ring. Here we
use the data obtained in July 2004 from the Cassini orbiter (image
PIA 06091 from the collection NASA/JPL/Space Science Institute).
This image consists of a stripe with the width of 29 pixels in the
radial direction. The natural length of the image is about $248$
kilometers, and the width is about $7.2$ kilometers. In such an
approximation we may neglect the curvature of the ring. The
function $f(t)$ obtained by the sample average is shown in
Fig.\ref{CWTfig}a.

As the signal is realvalued, we used the following pair of initial
conditions: $u(0,b) = f(b)$ and $v(0,b) = 0$. Owing to the bound
size, the Cauchy problem for the Eqs.
(\ref{systemU})--(\ref{systemV}) can be replaced by the boundary
problem of the first kind. For the real component of the function
we used the initial signal value. For the imaginary component zero
boundary conditions has been taken. Because it is necessary to get
the most good resolution for the present numeric sample, the based
frequency has been chosen as $\omega_0=\pi$. By the reason that
the inverse wavelet transform is not necessary, we neglect by the
deviation of the wavelet for the admissible conditions.

The results of numerical analysis, the modulus and the phase of
the wavelet-transform determined as
$$
|w(a,b)|=\sqrt{u^2(a,b)+v^2(a,b)}, \qquad \phi(a,b)=\arctan
\frac{u(a,b)}{v(a,b)},
$$
are shown in Fig.\ref{CWTfig}b,c. In this Figure, the larger
values correspond more dark regions.

Let us consider the modulus of the wavelet-transform
(Fig.\ref{CWTfig}b) in detail. One can easily see that there are
regions with monotonically changing instant periods. These regions
belong to the intervals $0.05-0.2$ and $0.5-0.85$. In the
non-regular region of the ring A periodic components do not
revealed. Note that to avoid the boundary effect, we do not
consider intervals closely related to the ends of the sample.

\begin{figure}
\begin{center}
\includegraphics[scale=0.5]{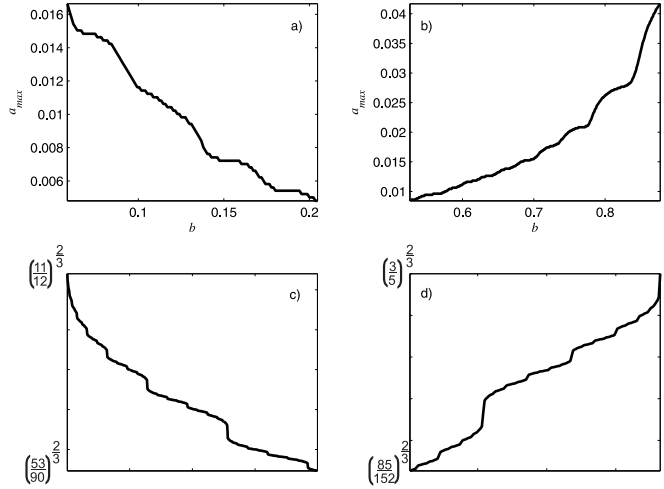}
\end{center}
\caption{Instant periods of the radial particle distribution in
the Saturn A-ring (a,b) and the Farey sequence (c,d).}
\label{comparis}
\end{figure}

As is known, values and dynamics of the instant period are
determined by the modulus maxima lines of the wavelet-transform.
Let us extract them from the region with the regular image (see
Fig.\ref{CWTfig}b). The obtained lines are shown in
Fig.\ref{comparis}a,b. The form of these lines confirm the
understanding of the fine stricture as the regions corresponding
to the high-order resonances with Saturn's moons. In particular,
the analyzed image PIA 06091 from the collection NASA/JPL/Space
Science Institute can be construed as the region between the
resonances $11:12$ Prometheus and $3:5$ Mimas. Using the third
Kepler law we may obtain the distribution of the major semiaxes
corresponding to the Farey sequence from this interval
(Fig.\ref{comparis}c,d). In this Figure, we clearly see the near
resemblance of the found parts of the sequence with the line of
the instant radial periods.

\section{The wavelet transform with the Morlet basis and
the correlation coefficient}

The rigorous definition of the Morlet basis (\ref{MorletExact})
makes it possible to find quantitatively the correlation of the
instant period with period of the monochromatic harmonics.
Introduce a new denotation $\alpha=a/\omega_0$ and rewrite the
corresponding integral convolution:
\begin{equation}
\label{FirstTerm}
 e^{-i\frac{b}{\alpha}}
  \int\limits_{ - \infty }^{ + \infty }
f(t)e^{i(\alpha^{-1})t}
 \frac{e^{
  -\frac{(t-b)^2}{2\alpha^2\omega^2_0}}}
{\sqrt{2\pi \alpha^2\omega^2_0}}dt.
\end{equation}
At $f(t)=1$ the integral (\ref{FirstTerm}) has a closed form
solution $e^{i\frac{b}{\alpha}}e^{-\frac{\omega_0^2}{2}}$. Thus,
the following representation is admissible:
\begin{equation}
\label{exp_omega} e^{-\frac{\omega_0^2}{2}}=
e^{-i\frac{b}{\alpha}}
  \int\limits_{ - \infty }^{ + \infty }
e^{i(\alpha^{-1})t}
 \frac{e^{
  -\frac{(t-b)^2}{2\alpha^2\omega^2_0}}}
{\sqrt{2\pi \alpha^2\omega^2_0}}dt.
\end{equation}
Therefore, the wavelet transform (\ref{CWT}) with the Morlet basis
(\ref{MorletExact}) can be rewritten as:
\begin{eqnarray*}
 w(a,b)=e^{-i\frac{b}{\alpha}}\left[
 \int\limits_{ - \infty }^{ + \infty }
f(t)e^{i(\alpha^{-1})t}
 \frac{e^{
  -\frac{(t-b)^2}{2\alpha^2\omega^2_0}}}
{\sqrt{2\pi \alpha^2\omega^2_0}}dt-\right.\\
 \left.\int\limits_{ - \infty }^{ +
\infty } e^{i(\alpha^{-1})t}
 \frac{e^{
  -\frac{(t-b)^2}{2\alpha^2\omega^2_0}}}
{\sqrt{2\pi \alpha^2\omega^2_0}}dt \int\limits_{ - \infty }^{ +
\infty } f(t)
 \frac{e^{
  -\frac{(t-b)^2}{2\alpha^2\omega^2_0}}}
{\sqrt{2\pi \alpha^2\omega^2_0}}dt \right]
\end{eqnarray*}
Let us compare here the expression in the square brackets with the
correlation coefficient of two random variables $x$ and $y$:
$$
R=\frac{\langle xy \rangle-\langle x \rangle \langle y
\rangle}{\sqrt{D_xD_y}},
$$
where $\langle\cdot\rangle$ means the averaging operation and
$D_\xi=\langle \xi^2 \rangle-\langle \xi \rangle^2$ is a
dispersion. We see that this wavelet transform is equivalent to
$$
w(\alpha,b)=e^{-i\frac{b}{\alpha}}\left[\langle
f(t)e^{i(\alpha^{-1})t} \rangle -\langle f(t) \rangle\langle
e^{i(\alpha^{-1})t} \rangle \right],
$$
and $t$ are normally distributed sample nodes. This distribution
is centered in $b$.

Let us introduce dispersions:
$$
D_f=\left|\int\limits_{ - \infty }^{ + \infty } f(t)^2
 \frac{e^{
  -\frac{(t-b)^2}{2\alpha^2\omega^2_0}}}
{\sqrt{2\pi \alpha^2\omega^2_0}}dt- \left( \int\limits_{ - \infty
}^{ + \infty } f(t)
 \frac{e^{
  -\frac{(t-b)^2}{2\alpha^2\omega^2_0}}}
{\sqrt{2\pi \alpha^2\omega^2_0}}dt\right)^2\right|,
$$
$$
D_{exp}=\left|\int\limits_{ - \infty }^{ + \infty }
e^{2i(\alpha^{-1})t}
 \frac{e^{
  -\frac{(t-b)^2}{2\alpha^2\omega^2_0}}}
{\sqrt{2\pi \alpha^2\omega^2_0}}dt- \left( \int\limits_{ - \infty
}^{ + \infty } e^{i(\alpha^{-1})t}
 \frac{e^{
  -\frac{(t-b)^2}{2\alpha^2\omega^2_0}}}
{\sqrt{2\pi \alpha^2\omega^2_0}}dt\right)^2\right|.
$$
The second expression has the following simple analytical
solution:
$$
D_{exp}=e^{-\omega_0^2}-e^{-2\omega_0^2}.
$$
Now we are able to find quantitatively the local correlation
coefficient of the signal $f(t)$ and the function
$\exp(i(\alpha^{-1})t)$ in a neighborhood of the given point
$t=b$:
\begin{equation}
\label{Corr}
    R_\alpha(b)=\frac{|w(\alpha,b)|}{\sqrt{{D_f}D_{exp}}}.
\end{equation}
All the integrals contained the analyzed function are the solution
of the Cauchy problem for the simplest diffusion equation:
$$
\frac{\partial u(\tau,b)}{\partial \tau}=\alpha^2\frac{\partial^2
u(\tau,b)}{\partial b^2}.
$$
These solutions are taken in time $\tau=\omega^2_0/2$. As an
initial condition we used $f(t)\exp(-(\alpha^{-1})t)$, $f(t)$ or
$f(t)^2$. To solve this equation there are robust methods (see,
e.g. \cite{Skeel}) even on the non-uniform grid.

\section{Conclusions}

Thus, we showed that continuous transform with the complex Morlet
wavelet is easily performed if we replace the integration of the
fast-oscillation function by the solution of the differential
equations. In spite of fact that this method requires more
computer resources, it has certain advantages. The most important
of them is that the initial data can be represented by non-uniform
sample of an arbitrary number nodes. In addition, this algorithm
is realized via the standard MATLAB. During the calculation, the
time variable in the diffusion equation is associated with the
scale variable of the wavelet transform. Therefore, the choice of
a small enough step allows us to trace the evolution of the
instant period in detail.

In the present article we apply the proposed method to the
processing of the image of the Saturn A-ring PIA 06091. This
analysis confirm the resonance origin in its structure.

More detail quantitative description is the subject of our
following analysis. This analysis is possible due to the video
material obtained from the Cassini spacecraft last half 2004 year.
These quantitative estimations can be based on results descried in
the second part of the article, where we got the dependence of the
local correlation coefficient of the signal and the harmonic
function in neighborhoods of the given point. Also, the
computation of this coefficient may be realized via the solution
of diffusion equations.

The proposed algorithm may be additionally considered as a certain
extension of the local time-series analysis. Namely, in
\cite{Loskutov}, a set of sequence with the compact support has
been advanced as test values. These sequences are obtained by the
periodization of sample's parts. In our analysis, as a test we
propose to use the harmonics.

\end{document}